\documentclass{eas}
\usepackage{graphicx}

\newcommand{\vsini}{v_{rot}\cdot\sin i}
\newcommand{\djdt}{\frac{dJ}{dt}}
\newcommand{\dmdt}{\frac{dM}{dt}}
\newcommand{\kms}{km~s$^{-1}$}
\begin{document}

\title{The role of magnetic fields in governing the angular momentum
  evolution of solar-type stars}

\runningtitle{The angular momentum evolution of solar-type stars}
\author{J. Bouvier}\address{Laboratoire d'Astrophysique, Observatoire
  de Grenoble, Universit\'e J. Fourier, CNRS, BP 53, 38041 Grenoble
  Cedex 9}
\begin{abstract}
I review the development of ideas regarding the angular momentum
evolution of solar-type stars, from the early 60's to the most recent
years. Magnetic fields are the central agent that dictates the
rotational evolution of solar-type stars, both during the pre-main
sequence, through star-disk magnetic coupling, and during the main
sequence, through magnetized winds. Key theoretical developments as
well as important observational results are summarized in this review.
\end{abstract}
\maketitle
\section{Introduction}

Magnetic fields hold clues to a number of properties of solar-type
stars. Among those, their rotation rate is found to vary along the
star's life, in response to a variety of physical processes, most of
which are intimately related to the stellar magnetic field. In this
review, I summarize what is currently known of the rotational
evolution of solar-type stars and emphasize the central role magnetic
fields play in governing angular momentum evolution. In Section~2, I
briefly review the main steps that have led to the current
understanding of rotational braking by magnetized winds on the main
sequence. In Section~3, I describe more recent advances related to the
angular momentum evolution of pre-main sequence stars and its
relationship to magnetic star-disk coupling in young stars. In
Section~4, these physical processes are illustrated through the
development of angular momentum evolution models that grasp the main
trends of the observed rotational evolution of solar-type stars, from
their birth up to the age of the Sun. I conclude by mentionning a few
theoretical and observational tracks that should provide an in-depth
understanding of these issues in the years to come.

\section {Magnetic braking on the main sequence }

In an illuminating review paper, Kraft (1970) summarized what was
known of the rotation rate of main sequence stars in the late
60's. One of the main results was a clear break in the measured
rotational velocities around a spectral type F5
(M$\sim$1.3~M$_\odot$), with a striking contrast between rapidly
rotating massive stars ($\vsini\sim 40-150$~\kms) and slowly rotating
solar-type stars ($\vsini \leq 25$~\kms). Schatzman (1959, 1962) had
anticipated this result by proposing a very efficient mechanism to
remove angular momentum from late-type, magnetically-active stars. He
suggested that the material ejected from the surface of solar-type
stars, mostly through magnetic flares, could carry away a large amount
of angular momentum, as the outflowing material is frozen in the
stellar magnetic field and thus kept in near-corotation with the star
up to large distances from the surface. The Weber \& Davis (1967)
solar-wind model indeed predicts an angular momentum loss rate that
scales with the square of the Alfv\'en radius,
$R_A$, \begin{equation}\djdt \propto\dmdt\cdot\Omega_\star\cdot
  R_A^2\label{eqdjdt}\end{equation} where $\dmdt$ is the mass loss
rate and $\Omega_\star$ the star's angular velocity, instead of
scaling merely as the square of the stellar radius, $R_\star$, in the
non-magnetized case. Since $\frac{R_A}{R_\star}\simeq30$ for the Sun,
the braking timescale by a magnetized wind is shorter by a factor of
$\sim$10$^3$ than what it is in the non-magnetized case.  Magnetized
stellar winds are thus quite efficient in braking late-type,
magnetically-active stars on a timescale of $\sim$10$^8$~yr during
their early evolution on the main sequence (Belcher \& MacGregor
1976).

Another major result outlined in Kraft's (1970) review, was the fact
that younger main sequence stars, as observed in young open clusters
at an age of $\sim$100~Myr, tend to rotate faster than mature field
dwarfs at an age of a few Gyr. Main sequence braking was further
quantified by Skumanich (1972), who showed from the rotation rates of
solar-type stars in the Pleiades, in the Hyades, and for the Sun, that
the rotational velocity decreases on the main sequence as the inverse
square-root of time, i.e., $\displaystyle v_{rot}(t)\propto
t^{-1/2}$. The so-called ``Skumanich law'' was readily interpreted in
the framework of Schatzman's theory. With an angular momentum loss
rate given by Eq.~\ref{eqdjdt}, and assuming the stellar magnetic flux
is produced by a linear dynamo, i.e., $R_\star^2B_\star\propto\Omega$,
one derives $\djdt \propto \Omega^3$ (Durney \& Latour 1978), which
integrates into $\Omega(t)\propto t^{-1/2}$. A general, albeit
parametric, expression for the angular momentum loss rate by a
magnetized wind has since then been worked out by Kawaler (1988).

As measurements of rotational velocities for solar-type stars on the
main sequence became more abundant, it became apparent that the
Skumanich relationship holds only on an average sense (Soderblom
1983). On the zero-age main sequence (ZAMS), at an age of about
100~Myr, solar-type stars actually exhibit a large intrinsic {\it
  dispersion} of rotational velocities, from less than 20~\kms\ up to
150~\kms\ (Stauffer et al. 1985). This result was surprising indeed,
as the extrapolation of the Skumanich relationship back in time from
the age of the Sun to the ZAMS would predict velocities of order of
20~\kms\ at 100~Myr. How this wide distribution of ZAMS rotation rates
builds up during pre-main sequence evolution will be discussed in the
next section.

Angular momentum evolution models were developed, using the observed
dispersion of rotational velocities on the ZAMS as initial
conditions. The braking rate was assumed to scale with velocity as
$\djdt\propto\Omega^3$, so that fast rotators on the ZAMS are more
efficiently spun down than slow ones during early MS evolution. As a
result, the large dispersion in initial velocities is quickly reduced
and converges towards uniformly slow rotation by the age of the Sun,
in qualitative agreement with observations. It soon became clear,
however, that a Skumanich type braking rate could not correctly
reproduce the evolution of the rotational distributions observed for
stars in open clusters of increasing age (cf. Fig.~1a). From empirical
modeling, MacGregor \& Brenner (1991) suggested instead that such a
braking rate applies to slow rotators only, and has a much shallower
dependence on rotation for rapid rotators, above some threshold
velocity, $\Omega_{sat}$.

Models using this alternative braking law were relatively successful
in reproducing the observations (cf. Fig.~1b). In these models, the
braking rate still scales as $\djdt\propto \Omega^3$ for slow rotators
($\Omega\leq\Omega_{sat}$), but only as $\djdt\propto \Omega$ for
rapid ones ($\Omega > \Omega_{sat}$). This is equivalent to assuming a
linear dynamo for slow rotators ($B\propto\Omega$) that saturates at
high rotation ($B\sim B_{sat}$). While the actual physical meaning of
``dynamo saturation'' remains somewhat unclear, it is nevertheless
supported by observations of chromospheric and coronal activity
diagnostics, whose strength saturates at velocities larger than
10-20~\kms\ in young solar-type stars (Vilhu 1984; Stauffer et
al. 1994). Furthermore, modeling the rotational evolution of
lower mass stars suggests that the saturation velocity,
$\Omega_{sat}$, depends on stellar mass, in such a way that saturation
occurs at a fixed Rossby number, $R_o=(\Omega \tau_{conv})^{-1}$,
where $\tau_{conv}$ is the turnover convective time, over the mass
range 0.5-1.0~M$_\odot$ (Collier Cameron \& Jianke 1994; Krishnamurthi
et al. 1997; Bouvier et al. 1997). This empirical result tends to
provide some theoretical support to the concept of dynamo saturation
in low-mass, partly convective stars.

   \begin{figure}
   \centering
        \resizebox{0.50\textwidth}{!}{\includegraphics  {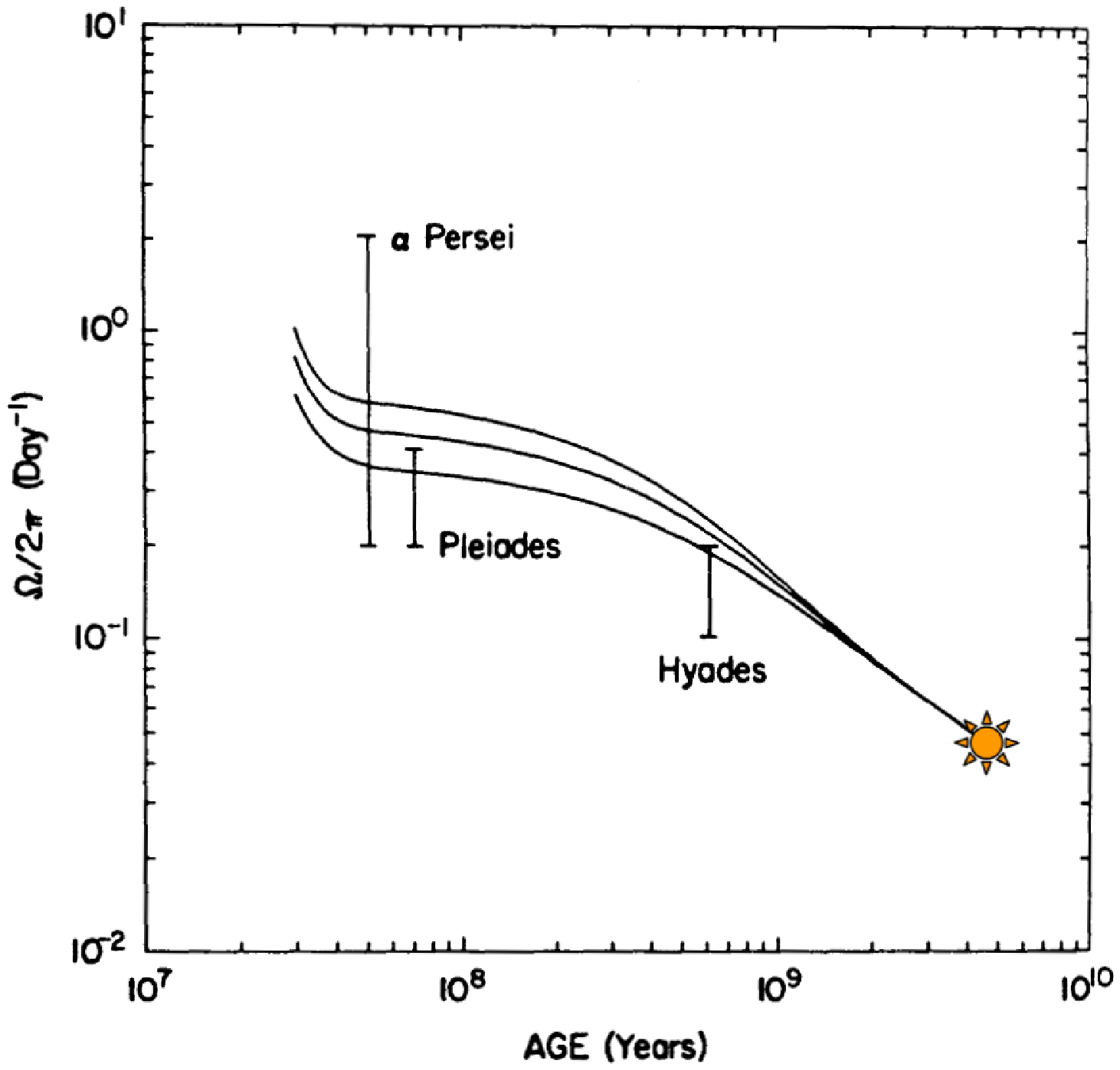}}
        \resizebox{0.47\textwidth}{!}{\includegraphics  {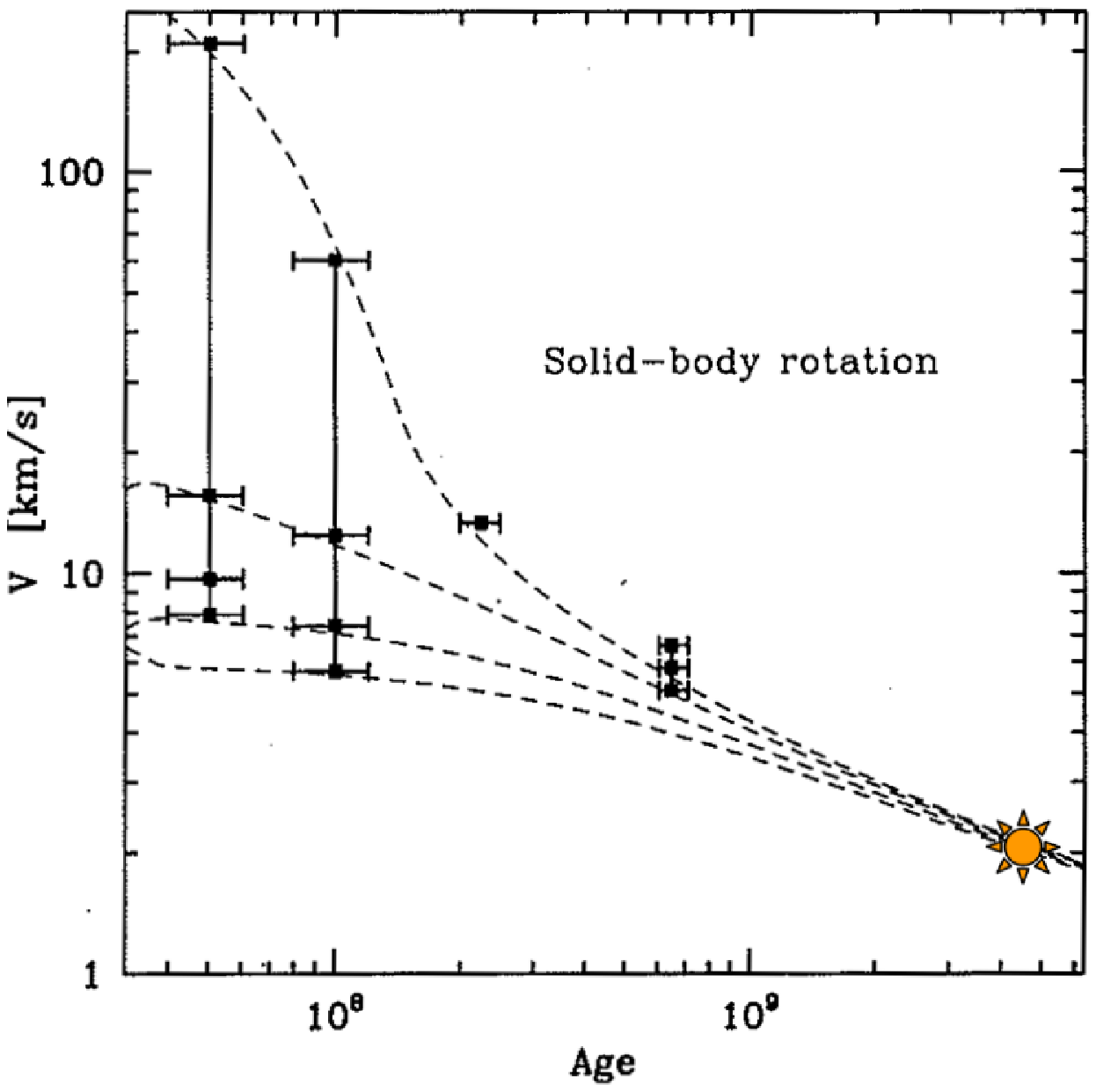}}
   \caption{{\it Left~:} Angular momentum evolution models from the
     ZAMS to the age of the Sun, assuming a braking rate $ \djdt
     \propto \Omega^3$. The range of velocities observed for
     solar-type stars in 3 young open clusters is shown as
     bars. Models are shown for 3 initial velocities on the
     ZAMS. Adapted from MacGregor \& Brenner (1991).  {\it Right~:}
     Same as left panel but for models assuming dynamo saturation at
     $\Omega_{sat}\sim 14~\Omega_\odot$ (see text). Models are shown
     for 4 initial velocities on the ZAMS. The rotational
     distributions of solar-type stars in young open clusters are
     shown by vertical bars with 10, 20, 50 and 90 percentiles
     indicated. The remaining discrepancies between models and
     observations, most notably for slow rotators, arise from the
     assumption of solid-body rotation (see text). Adapted from
     Bouvier (1997). }
    \end{figure}

Finally, the response of surface velocity to angular momentum loss at
the stellar surface also depends on how efficiently angular momentum
is transported in the stellar interior. A rigidly-rotating star has a
longer spindown timescale than a differentially-rotating one in which
only the outer convective envelope is braked while the radiative core
remains in rapid rotation. This is because the moment of inertia of
the convective envelope is much smaller than that of the whole
star. Core-envelope decoupling (Endal \& Sofia 1981) thus enters as an
additional parameter in angular momentum evolution models. A physical
description of angular momentum transport in stellar interiors was
developed by the Yale group, with some success (Pinsonneault et
al. 1989). Their rotational models, however, predicted a
rapidly-rotating core for the present-day Sun, in conflict with
heliosismological results. A more empirical approach was designed by
MacGregor \& Brenner (1991), who assumed that both the radiative core
and the convective envelope rotate uniformly, but not necessarily at
the same rate ($\Omega_{core} \neq \Omega_{conv}$). They introduced a
coupling timescale between the core and the envelope, $\tau_c$, which
measures the time it would take to transfer angular momentum from the
core to the envelope in order to restore uniform rotation throughout
the star. A short coupling timescale enforces rigid-body rotation
between the core and the envelope, while a long one leads to the
development of a large velocity shear at the core-envelope
boundary. Recent models based on updated observational constraints
suggest that the coupling timescale itself depends on rotation, being
much longer in slow rotators ($\sim$100~Myr) than in fast ones
($\sim$10~Myr) (Bouvier 2008, see Sect.~4).

Current models of angular momentum evolution for solar-type (and
lower mass) stars on the main sequence thus rely on~: i) initial
conditions derived from measurements of surface rotation for ZAMS
stars in young open clusters, ii) a parametrized braking law that
includes dynamo saturation at high velocity, and iii) core-envelope
decoupling. Unfortunately, no complete quantitative theory exists yet
for angular momentum loss due to magnetized stellar winds, nor for
angular momentum transport in stellar interiors. As a result, the
models include these physical processes through simplified,
parametrized relationships, with parameters being the normalisation of
the magnetic braking rate (usually scaled on the Sun's), the velocity
at which dynamo saturation occurs (somewhat constrained by the
observations of magnetic activity diagnostics), and the timescale for
angular momentum exchange between the core and the envelope (only
loosely constrained by the Sun's internal rotation profile). The
impact the various parameters have onto rotational evolutionary tracks
is illustrated in Keppens et al. (1995), Bouvier et al. (1997) and
Allain (1998). Combining these ingredients, recent models are
reasonably successful in reproducing the observed evolution of the
rotational distributions of solar-type and lower mass stars on the
main sequence (e.g. Irwin et al. 2008).

\section {Magnetic star-disk coupling during the pre-main sequence}

Since internal differential rotation plays a key role in the evolution
of the surface rotation rate of solar-type stars, and because the
amount of core-envelope decoupling in ZAMS stars is unknown, initial
conditions for angular momentum evolution models have to be set at the
very start of pre-main sequence (PMS) evolution, when the stars are
still fully convective and may thus be assumed to be in solid-body
rotation. The derivation of statistically robust rotational
distributions for low-mass PMS stars at an age of $\sim$1~Myr, the
so-called T Tauri stars (TTS), has been the goal of numerous studies
over the past 20 years.
 
While the extrapolation of the Skumanich law back in time from the
Sun's age to 1~Myr predicted that TTS should have velocities as high
as 150~\kms, early measurements showed instead that they were
relatively slow rotators, with $\vsini$ values of order of 20~\kms\ or
less (Vogel \& Kuhi 1981; Bouvier et al. 1986; Hartmann et
al. 1986). Hartmann \& Stauffer (1989) further pointed out that young
stars accreting high angular momentum material from their
circumstellar disk ought to spin up at velocities close to break-up on
a timescale of order of 1~Myr. Obviously, some extremely efficient
braking mechanism must be at work in these stars in order to prevent
them from spinning up during their early PMS evolution.

Following a model originally proposed by Ghosh \& Lamb (1979) in the
context of accreting neutron stars, Camenzind (1990) and K\"onigl
(1991) suggested that the magnetic coupling between the young active
star and its inner accretion disk could lead to angular momentum
transport that would effectively extract angular momentum from the
central star, thus braking it. A key observational result in support
to this hypothesis was that accreting TTS appeared to have, on
average, lower rotation rates than non-accreting ones (Edwards et
al. 1993; Bouvier et al. 1993). Somewhat paradoxically, it thus seemed
that slow rotation was intimately linked to disk accretion in young
stars.

Since then, evidence has been growing in support of magnetic star-disk
coupling in young stars (Alencar 2007; Bouvier et al. 2007a, 2007b)
and a number of theoretical and numerical studies have investigated
how such a magnetic coupling could brake the young star much more
efficiently than a solar-type magnetized wind would (e.g. Shu et
al. 2007; Romanova et al. 2007; Matt \& Pudritz 2008). Additional
evidence has been reported in the Spitzer era for slower rotation
rates in accreting TTS (Rebull et al. 2006; Cieza \& Baliber 2007). As
early envisionned by Choi \& Herbst (1996), the rotational period
distribution of young stars appears bimodal, consisting of a peak of
slow rotators at $P\sim 8$~days, and a wide tail of faster
rotators. Slow rotators with infrared excess are assumed to be stars
magnetically locked to their disk, and thus prevented from spinning
up. Fast rotators without infrared excess have already dissipated
their disk and have had enough time to spin up as they contract
towards the main sequence.

Then again, albeit in quite a different way, magnetic fields, and more
specifically the magnetic star-disk interaction, seem to dictate the
angular momentum evolution of young stars during the
PMS. Observationally, the lifetime of circumstellar disks is found to
vary from star to star, over the range $\sim$1-10~Myr (Hillenbrand
2005; Meyer et al. 2007). Stars with long-lived disks are prevented
from spinning up for a significant fraction of their PMS evolution and
consequently reach the ZAMS as slow rotators. On the opposite, stars
with short-lived disks freely spin up as they contract towards the
ZAMS, which they reach as fast rotators as solar-type magnetic winds
are unable to brake them on such short timescales (cf. Matt \& Pudritz
2007). Thus, the wide dispersion of rotational velocities observed on
the ZAMS for solar-type stars ($\leq$20-150~\kms, see Sect.~2) largely
results from the distribution of disk lifetimes in the PMS (Bouvier et
al. 1993; Bouvier 1994; Collier Cameron et al. 1995).

\section {Angular momentum evolution models}

   \begin{figure}
   \centering
  \includegraphics[width=1.0\textwidth]{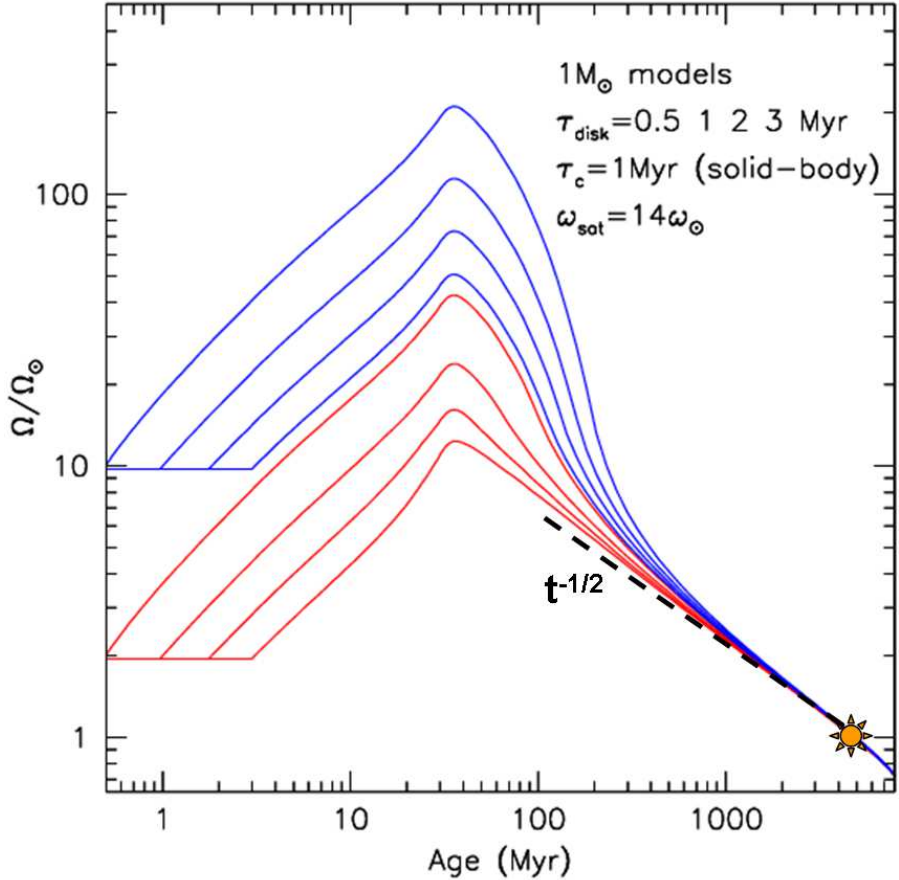}
   \caption{Grid of rotational models for 1~M$_\odot$ stars. Starting
     from 2 initial velocities, rotational tracks are computed for
     disk lifetimes of 0.5, 1, 2, and 3~Myr (from left to
     right). These models assume a core-envelope coupling timescale of
     1~Myr, which is equivalent to solid-body rotation. The velocity
     at which dynamo saturation occurs is taken to be
     14~$\Omega_\odot$. Note the convergence towards uniformly slow
     rotation at the Sun's age, regardless of the past rotational
     history. The t$^{-1/2}$ Skumanich relationship is shown for
     reference. The 1~M$_\odot$ structural evolution model is from
     Baraffe et al. (1998). }
              \label{grid}%
    \end{figure}

Current angular momentum evolution models include the various physical
mechanisms described in previous sections, albeit usually in a
parametrized and/or phenomenological way. A grid of rotational models
is shown for solar-mass stars in Fig.~\ref{grid}. The main phases of
the evolution of surface rotation are clearly seen. As long as the
star is coupled to its disk during the early PMS, it is assumed to
evolve at a constant angular velocity. Once the disk has dissipated, a
saturated braking law is applied, but solar-type winds are inefficient
to prevent the star from spinning up as it contracts towards the
ZAMS. Once on the ZAMS, the structure of the star stabilizes and
magnetic braking becomes dominant. The assumed braking law leads to a
convergence towards slow rotation within a few 100~Myr on the main
sequence. Indeed, the surface rotation rate of mature solar-type
stars has lost memory of the past rotational history.

\begin{figure}
   \centering
  \includegraphics[width=1.0\textwidth]{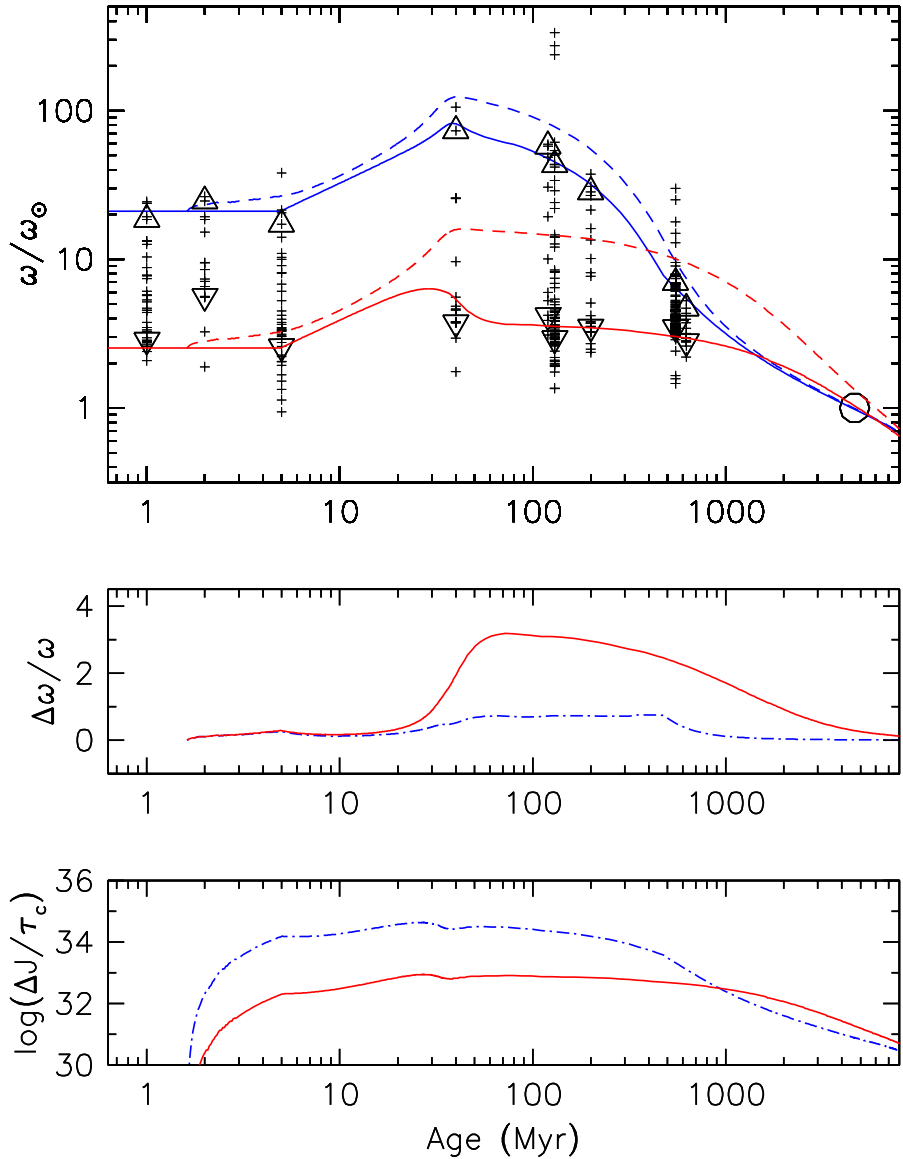}
   \caption{Rotational models for slow and fast solar-mass
     rotators. {\it Upper panel~: observations and models.} The 10th
     and 75th percentiles of the observed rotational period
     distributions of solar-type stars (0.8-1.1~M$_\odot$) were
     converted to angular velocity and are plotted as direct and
     inverted triangles as a function of time. Individual measurements
     of rotational periods converted to angular velocities are also
     shown in order to illustrate the statistical significance of the
     various samples. The modeled evolution of surface rotation for
     slow and fast rotators is shown by the solid lines. For both
     models, the rotation of the radiative core is shown by the dashed
     lines. With a core-envelope coupling timescale of only 10~Myr,
     little differential rotation develops in fast rotators. In
     contrast, the 100~Myr core-envelope coupling timescale in slow
     rotators leads to the development of a large velocity gradient at
     the base of the convective zone. A disk lifetime of 5~Myr is
     assumed for both models. {\it Lower panels :} The velocity shear
     at the base of the convective zone,
     ($\omega_{rad}-\omega_{conv})/\omega_{conv}$, and the angular
     momentum transport rate, $\Delta J/\tau_c$ ($g cm^2 s^{-2}$),
     from the core to the envelope are shown for slow (solid line) and
     fast (dotted-dashed line) rotators. From Bouvier (2008). }
              \label{model}%
    \end{figure}

These models are confronted to observations in Figure~\ref{model}.  In
the last years, rotational periods have been measured for hundreds of
low-mass stars in molecular clouds (1-5~Myr) and young open clusters
(40-600~Myr), thus providing a tight observational sampling of the
rotational evolution of low-mass stars (e.g. Irwin et
al. 2008). Figure~\ref{model} shows these rotational period
distributions, converted to angular velocity ($\omega = 2\pi/P$), for
solar-type stars in a number of young and intermediate-age clusters
(see Bouvier 2008 for complete references on the datasets). The 2
models shown in Fig.~\ref{model} illustrate the evolution of the
slowest and fastest rotators. For the sake of clarity, models for
intermediate rotators are not shown, but would have a similar
behaviour, with parameters intermediate between those of slow and fast
rotator models.

The model for fast rotators starts from an initial period of
1.2~d. The star is assumed to remain coupled to its disk for 5~Myr,
then spins up to a velocity of order of 160~km~s$^{-1}$ on the ZAMS,
and is eventually spun down by a magnetized wind on the MS to the
Sun's velocity. This model fits reasonably well the PMS spin up and
the rapid MS spin down observed for fast rotators between 5 and
500~Myr. In order to reach such an agreement, the core-envelope
coupling timescale has to be short, $\tau_c \sim $ 10~Myr. A longer
coupling timescale would lead to envelope spin down before the star
reaches the ZAMS, and a slower spin down rate on the early MS, both of
which would conflict with observations. The tight coupling between the
core and the envelope implies that little differential rotation
develops in fast rotators, with the rotation of the core barely
exceeding that of the envelope on the early MS (see Fig.~\ref{model}).

The slow rotator model has an initial period of 10~d and the star-disk
magnetic interaction is assumed to last for 5~Myr in the early PMS. As
the star approaches the ZAMS, both the outer convective envelope and
the inner radiative core spin up. Once on the ZAMS, however, only the
outer envelope is quickly braked, while the core remains in rapid
rotation. This behaviour results from an assumed weak coupling between
the core and the envelope, with $\tau_c\sim$ 100~Myr. On the early MS,
the rapidly-rotating core transfers angular momentum back to the
envelope, which explains the nearly constant surface velocity over
several 100~Myr in spite of magnetic braking. A long coupling
timescale between the core and the envelope is thus required to
account for the observed rotational evolution of slow rotators. A long
$\tau_c$ in slow rotators implies inefficient transport of internal
angular momentum and results in a large velocity gradient at the
core-envelope boundary (see Fig.~\ref{model}).

The main difference between fast and slow rotators in these models is
thus twofold~: the initial angular momentum and the level of
core-envelope decoupling. The initial angular momentum of fast
rotators is 10 times higher than that of slow ones. Fast rotators lose
more angular momentum over the course of their evolution than slow
rotators, as they both converge towards the Sun's velocity at
4.65~Gyr. The lower panels in Fig.~\ref{model} show the magnitude of
the rotational shear at the base of the convective envelope for slow
and fast rotators, and the amount of angular momentum transported from
the core to the envelope. Strong differential rotation develops in
slow rotators at the ZAMS and remains large during early MS evolution
until the core and the envelope eventually recouple after a few
Gyr. Clearly, slow rotators exhibit a much larger rotational shear at
the base of their convective envelope than fast rotators during most
of their evolution. Conversely, angular momentum transport from the
core to the envelope is much more efficient in fast rotators than in
slow ones, which results in little differential rotation indeed.
Rotational shear and angular momentum transport are both directly
related to the assumed core-envelope coupling timescale and are
independent of the disk lifetime. The implications of the different
internal rotation profiles between slow and fast rotators for the
properties of mature solar-type stars and for the planet formation
process in circumstellar disks have been discussed in Bouvier (2008).

\section {Conclusions} 

Stellar magnetic fields are the central ingredient governing the
rotational evolution of solar-type stars. Star-disk magnetic coupling
during pre-main sequence evolution dictates the rotational history of
young suns, up to their arrival on the zero-age main
sequence. Magnetized stellar winds then become dominant on the main
sequence and yield uniformly slow rotation by the age of the
Sun. Semi-empirical models are relatively successful in reproducing
the main trends of the observed rotational evolution of solar-type
stars from their birth to the Sun's age. Yet, these models merely
include phenomenological descriptions of the physical processes at
work. An area of progress for the years to come will hopefully be the
development of complete physical theories for the star-disk magnetic
interaction, for angular momentum loss through magnetized winds, and
for instabilities that redistribute angular momentum in stellar
interiors. On the observational side, spectropolarimetric measurements
will provide unprecedented details on the strength and topology of
stellar magnetic fields in solar-type stars at various ages (see
Donati, this volume). Such observations will shed light on the nature
of the star-disk interaction in young stars (Donati et al. 2008),
provide physical clues to the empirical concept of dynamo saturation
(Petit et al. 2008), and help extrapolating the angular momentum
evolution models developed for solar-type stars to lower mass stars
(Morin et al. 2008).


\end{document}